# Nutmeg and SPICE: Models and Data for Biomolecular Machine Learning

Peter Eastman[1], Benjamin P. Pritchard[2], John D. Chodera[3], Thomas E. Markland[1]

[1]Department of Chemistry, Stanford University, Stanford, CA 94305, USA
[2]Molecular Sciences Software Institute, Virginia Polytechnic Institute and State University, Blacksburg, VA 24060, USA
[3]Computational and Systems Biology Program, Sloan Kettering Institute, Memorial Sloan Kettering Cancer Center, New York, NY 10065, USA

corresponding author: Peter Eastman (peastman@stanford.edu)

# Abstract

We describe version 2 of the SPICE dataset, a collection of quantum chemistry calculations for training machine learning potentials. It expands on the original dataset by adding much more sampling of chemical space and more data on non-covalent interactions. We train a set of potential energy functions called Nutmeg on it. They are based on the TensorNet architecture. They use a novel mechanism to improve performance on charged and polar molecules, injecting precomputed partial charges into the model to provide a reference for the large scale charge distribution. Evaluation of the new models shows they do an excellent job of reproducing energy differences between conformations, even on highly charged molecules or ones that are significantly larger than the molecules in the training set. They also produce stable molecular dynamics trajectories, and are fast enough to be useful for routine simulation of small molecules.

# Introduction

Machine learning potentials are a popular tool for molecular simulation.[1] A large model, typically a neural network, is trained on a large body of forces and/or energies computed with a high level quantum chemistry method. The model learns to compute forces and energies for new conformations and, in some cases, new molecules. They represent a middle ground between conventional force fields, which are faster but have limited accuracy and transferability, and quantum chemistry methods, which are accurate but very slow.

The two essential ingredients that make up a machine learning potential are the model architecture and the data it is trained on. Much work has recently been done to develop both of these ingredients. Many flexible, general purpose architectures have been designed that can be applied to broad areas of chemical and conformational space.[2–5]

Many quantum chemistry datasets have also been introduced, but their generality tends to be much lower. A model is unlikely to work for situations far outside the domain of the data it was trained on. Given that chemical and conformational space are nearly infinite, any dataset will necessarily be limited in the applications for which it can be used. For example, some of the most popular datasets contain only a very small number of chemical elements.[6–8] Some contain only energy minimized conformations,[6,9,10] which limits their usefulness for training potential functions that can be applied to molecular dynamics. Some contain only neutral molecules,[7,9] limiting their usefulness for learning to simulate charged molecules. There is thus an ongoing need for more datasets that can be used for other applications.

SPICE is a quantum chemistry dataset designed for training machine learning potentials.[11] Its focus is particularly on modeling drug-like small molecules interacting with proteins. The original version contains forces and energies for 1.1 million conformations of molecules covering a wide range of chemical space, including drug molecules, dipeptides, and solvated amino acids. There are 15 elements in total, charged and uncharged molecules, low and high energy conformations, and a wide variety of covalent and non-covalent interactions.

In this article we describe an updated version of the SPICE dataset. It greatly expands the coverage of chemical space with over 20,000 new molecules. It also improves sampling of non-covalent interactions and adds two more elements (boron and silicon). In total, the new version 2 of SPICE contains approximately twice as much data as version 1.

Having created the new dataset, we train a set of machine learning potentials called Nutmeg on it. They are based on the TensorNet architecture.[4] They use a novel method to improve accuracy on charged and polar molecules. Static, pre-computed atomic partial charges are provided to the model as inputs, giving it detailed information about the average charge distribution. They also incorporate a short ranged repulsive term to alleviate instabilities encountered in many machine learning potentials. We validate the models on a variety of tasks, demonstrating their ability to produce accurate energies and stable simulation trajectories.

# Methods

## SPICE 2

SPICE was designed to be a living dataset. It will grow with time to allow training new models that are more accurate and transferable. Periodic versioned updates will allow for reproducibility. SPICE 2 is the first major update.

SPICE is composed of a collection of data subsets. Each subset is designed to provide a certain type of information, such as certain types of interactions or classes of molecules. The subsets are not intended to be used individually. The goal is that when all subsets are combined, they collectively provide enough information to train models that accurately describe broad areas of chemical space.

SPICE 2 keeps all the subsets from SPICE 1 (although a small number of calculations were rerun as described below). It adds the following new data subsets. For full details, the scripts used to generate them are available at https://github.com/openmm/spice-dataset.

### More PubChem Molecules

SPICE 1 contained 14,644 molecules with up to 50 atoms drawn from PubChem. SPICE 2 adds 9913 more PubChem molecules, chosen through the same procedure as the ones in SPICE 1. 50 conformations (25 high energy, 25 low energy) were generated for each one as follows.

RDKit 2023.03.1[12] was used to generate 10 conformations for each molecule. Starting from each one, 100 ps of molecular dynamics at 500K was performed using OpenMM 8.0[13] with the OpenFF 2.0.0 force field.[14] A conformation was saved every 10 ps, giving 100 total conformations. 25 of these that were maximally different from each other as measured by all atom RMSD were kept as the 25 high energy conformations. Starting from each of those, five iterations of L-BFGS energy minimization were performed, followed by 1 ps of molecular dynamics at 100K. These were kept as the 25 low energy conformations.

### PubChem Molecules with Boron and Silicon

Molecules containing B and Si were originally omitted, both because of the infrequency of these elements and because they are not supported by the OpenFF force field. SPICE 2 includes a new subset for PubChem molecules that pass all the other requirements used for selecting molecules, and contain one of these elements. In total there are 1562 molecules containing B and 1952 molecules containing Si. 50 conformations were generated for each one as described above, except using the GFN-FF force field.[15]

### Ligand, Amino Acid Pairs

This subset is designed to improve sampling of non-covalent protein-ligand interactions. Every molecule in Ligand Expo[16] meeting the following requirements was selected.

- It is not a standard amino acid, nucleotide, or water.
- It contains only the following elements: H, C, N, O, F, P, S, Cl, Br, I.
- It contains no more than 36 atoms, including hydrogens.
- It has no radical electrons.
- RDKit does not produce any sanitization errors while processing it.
- It can be parametrized with the OpenFF 2.0.0 force field.

A total of 10,584 molecules satisfied these requirements. For each one, a single PDB structure containing the ligand was selected. Every amino acid within 4 Å of the ligand was identified. Each amino acid was capped with ACE and NME groups. A conformation was generated by energy minimizing the ligand and a single amino acid together while applying harmonic restraints with force constant 1000 kJ/mol/nm$^2$ to all heavy atoms in the ligand and amino acid. This produced a total of 194,174 conformations.

**Solvated PubChem Molecules**

This subset is designed to sample non-covalent ligand-water interactions. 1397 PubChem molecules were simulated in a 2.2 nm cubic box of water at 300K for 200 ps using the OpenFF 2.0.0 force field and TIP3P-FB[17] water model. A conformation was saved every 20 ps consisting of the PubChem molecule and the 20 water molecules closest to it.

**Water Clusters**

To further improve the sampling of interactions in bulk water, 1000 conformations each consisting of 30 water molecules were generated. A 2 nm box of AMOEBA[18] water molecules was simulated at 300K for 10 ns. Every 10 ps, the 30 water molecules closest to the center of the box were saved as a conformation.

**Rerun Calculations from SPICE 1**

A small fraction of the DFT calculations run for version 1 of SPICE were found to be poorly converged due to a bug in the version of Psi4 used to perform them. This led to inaccurate energies and very large forces. When creating the previous release, we excluded these calculations by applying a filter based on force magnitude. These calculations have been rerun as part of version 2, allowing them to be included.

**Summary of SPICE 2 contents**

In total SPICE 2 contains 2 million conformations. The content is summarized in Table 1. All calculations used the ωB97M-D3(BJ) DFT functional[19,20] and def2-TZVPPD basis set.[21,22] They were run with Psi4[23] version 1.8.1 or 1.8.2. The workflow for running calculations was managed with QCArchive.[24]

| Subset | Molecules | Conformations | Atoms | Elements |
|---|---|---|---|---|
| Dipeptides | 677 | 33,850 | 26–60 | H, C, N, O, S |
| Solvated Amino Acids | 26 | 1300 | 79–96 | H, C, N, O, S |
| DES370K Dimers | 3490 | 345,676 | 2–34 | H, Li, C, N, O, F, Na, Mg, P, S, Cl, K, Ca, Br, I |
| DES370K Monomers | 374 | 18,700 | 3–22 | H, C, N, O, F, P, S, Cl, Br, I |
| PubChem | 28,039 | 1,398,566 | 3–50 | H, B, C, N, O, F, Si, P, S, Cl, Br, I |
| Solvated PubChem | 1397 | 13,934 | 63–110 | H, C, N, O, F, P, S, Cl, Br, I |
| Amino Acid Ligand Pairs | 79,967 | 194,174 | 24–72 | H, C, N, O, F, P, S, Cl, Br, I |
| Ion Pairs | 28 | 1426 | 2 | Li, F, Na, Cl, K, Br, I |
| Water Clusters | 1 | 1000 | 90 | H, O |
| Total | 113,999 | 2,008,628 | 2–110 | H, Li, B, C, N, O, F, Na, Mg, Si, P, S, Cl, K, Ca, Br, I |

**Table 1.** The overall content of the SPICE 2 dataset. Rows correspond to the subsets that make it up. Columns indicate 1) the number of distinct molecules or clusters in the subset, 2) the total number of conformations, 3) the minimum and maximum number of atoms of the molecules or clusters it contains, and 4) the list of elements found in the molecules it contains.

# Models

A set of machine learning potentials was trained on the SPICE 2 dataset. The goal of the models is to simulate molecules that cover the full range of the training data: drug-like small molecules and peptides, including charged and polar molecules, in all conformations likely to be encountered in a room temperature molecular dynamics simulation. They are not intended to simulate chemical reactions. The SPICE dataset does not include conformations with partially formed bonds, and a model trained on it is unlikely to provide a realistic description of the process of forming or breaking bonds.

All models used the TensorNet[4] architecture as implemented in PhysicsML.[25] This is an equivariant message-passing architecture. As shown in Figure 1, it consists of the following modules: an embedding layer, a configurable number of interaction layers, and an output layer. The embedding layer and each interaction layer involves one round of message passing. This allows each atom to exchange information with all other atoms that are within a cutoff distance $r_c$. A model with $n$ interaction layers therefore involves $n$+1 rounds of message passing and has a total receptive field (the maximum distance over which atoms can interact) of $(n+1)r_c$. Especially when simulating charged molecules, having a large receptive field is important if the model is to produce accurate results for large systems.

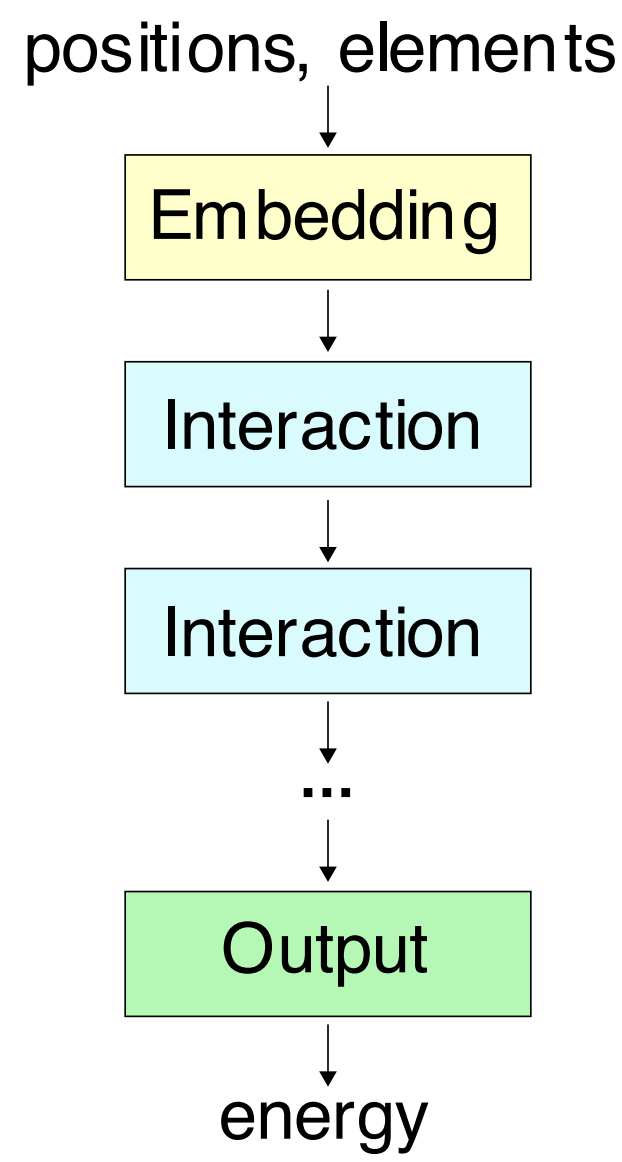

**Figure 1.** The TensorNet model architecture. It consists of an embedding layer to produce a vector describing each atom, followed by a configurable number of interaction layers, and finally an output layer to produce energy or other outputs.

Three models of varying sizes were trained. The smallest one has a single interaction layer of width 256. The medium model increases the width to 512, and the large model adds a second interaction layer. We refer to the models as Nutmeg-small, Nutmeg-medium, and Nutmeg-large. Detailed hyperparameters for all three models are shown in Table 2. They were selected through a process of manual testing to identify the factors with the largest impacts on both speed and accuracy.

| | **Small** | **Medium** | **Large** |
|---|---|---|---|
| Interaction layers | 1 | 1 | 2 |
| Width | 256 | 512 | 512 |
| Embedding width | 128 | 128 | 128 |
| Radial basis functions | 64 | 64 | 64 |
| Cutoff (nm) | 1.0 | 1.0 | 1.0 |
| Total parameters (millions) | 1.7 | 4.6 | 7.1 |

**Table 2.** Hyperparameters for the three Nutmeg models.

# Charge Distribution

It is an open question how machine learning potentials should handle the redistribution of charge in a molecule. The simplest approach is just to ignore it and hope the model will successfully infer the charge distribution based on the atom positions.[26] Unfortunately, this tends not to work well for charged molecules. It also is impossible, in general, to determine the charge distribution only from a local calculation, since Coulomb interactions are long ranged and charge redistribution can happen over long distances.

Some models use a global calculation to predict atomic charges, then compute an explicit Coulomb energy based on them.[27,28] This has the potential to accurately model large scale charge redistribution, but it is complicated to implement and expensive to compute. There also is not yet agreement on the best way to compute the charges.

A middle ground between these approaches is to precompute a fixed charge distribution independent of conformation. Conventional force fields produce surprisingly accurate results using fixed partial charges. Even better accuracy is obtained with polarizable force fields that allow charge to move over very short distances, such as with Drude particles or induced atomic dipoles.[29] We therefore hypothesize that a machine learning potential can produce accurate results on charged molecules if it is provided with a static description of the large-scale charge distribution to use as a baseline.

In previous work[11], we did this by incorporating formal charges into the atom typing system: the type of each atom was specified by the combination of element and formal charge. This simple approach substantially improved the accuracy of models for charged and polar molecules. Formal charges are only a very approximate description of the charge distribution, however. For the current work, we instead precompute Gasteiger partial charges[30] for atoms, which are used as an input to the model. This is a much more detailed description of the charge distribution, but is still independent of conformation and needs to be computed only once at the start of a simulation.

The first step in a TensorNet model is to multiply a feature vector for each atom by a linear transform to produce an initial embedding vector. If the feature vector is a one-hot encoded atomic number, the columns of the transform matrix are the embedding vectors for different elements. We generalize this by appending each atom's partial charge to the end of the feature vector, leading to an embedding vector that mixes together information about the element and charge in a learnable way.

We emphasize that the charge is injected into the model in a completely generic way. Unlike some other architectures, there is no explicit Coulomb interaction. It is up to the model to learn how to use partial charges most effectively.

A limitation of this method is that it cannot be used when simulating chemical reactions. The partial charges depend on which atoms are bonded to each other. Any change to the covalent bonding pattern would require the charges to be recomputed. This is not a significant limitation in the current case: the SPICE dataset contains no data for conformations with partially formed chemical bonds, so a model trained on it is unlikely to be able to break or form bonds in a realistic way. Our goal is to simulate conformational changes of systems consisting of whole molecules that cannot break apart or reform.

## Short Range Repulsion

A common problem in machine learning potentials is the lack of training data for physically inaccessible regions of configuration space. If two atoms are directly on top of each other, the energy should be extremely high, and therefore this situation should never occur in simulations. Because the training set contains no conformations in which atoms are on top of each other, however, the model has no opportunity to learn this. It could just as easily produce a very low energy, leading to unrealistic and unstable situations.

To avoid this problem, we add an explicit repulsive potential at short distances. We choose the Ziegler-Biersack-Littmark (ZBL) potential[31], an empirical potential that accurately describes the screened nuclear repulsion between overlapping atoms. It was parameterized to describe the scattering of ions from the atomic nuclei in solids, and gives an excellent fit for a wide variety of elements. We add the following energy for each pair of atoms separated by a distance *r*.

$$E(r) = \psi(r) \cdot ZBL(r)$$

ZBL(*r*) is the ZBL potential,

$$ZBL(r) = \frac{1}{4\pi\epsilon_0}\frac{Z_1 Z_2}{r}\phi(r/a)$$

$$a = \frac{0.8854a_0}{Z_1^{0.23} + Z_2^{0.23}}$$

$$\phi(x) = 0.1818e^{-3.2x} + 0.5099e^{-0.9423x} + 0.2802e^{-0.4029x} + 0.02817e^{-0.2016x}$$

where $Z_1$ and $Z_2$ are the atomic numbers of the interacting atoms and $a_0$ is the Bohr atomic radius. The ZBL potential is only applicable to atoms that are very close together. It does not provide an accurate description of covalently bonded atoms. We therefore restrict it to short distances by multiplying by the cutoff function

$$\psi(r) = \begin{cases} 0.5(1 + \cos(\pi r/r_{max})) & r < r_{max} \\ 0 & r \geq r_{max} \end{cases}$$

where the cutoff distance $r_{max}$ equals the sum of the covalent radii (taken from Slater[32]) of the two interacting atoms.

## Model Training

The models were trained on the SPICE 2 dataset with PhysicsML on four NVIDIA H100 GPUs. 5% of the conformations in the dataset were randomly selected and reserved as a validation set (100,251 conformations), and the rest were used as the training set. An AdamW optimizer was used with batch size 64. The loss function was a weighted sum of the L2 losses for energies and forces, with a weight of 1 $(kJ/mol)^{-2}$ for the energy loss and 0.01 $(kJ/mol/nm)^{-2}$ for the force loss. The learning rate was initially set to $5 \cdot 10^{-4}$, and reduced by multiplying by 0.8 after any epoch in which the validation loss failed to decrease, down to a minimum learning rate to $10^{-5}$. All models were trained for 100 epochs, by which point they were very well converged as shown in Figure 2. The validation set loss was computed after each epoch, and the version with the lowest loss was saved as the final model.

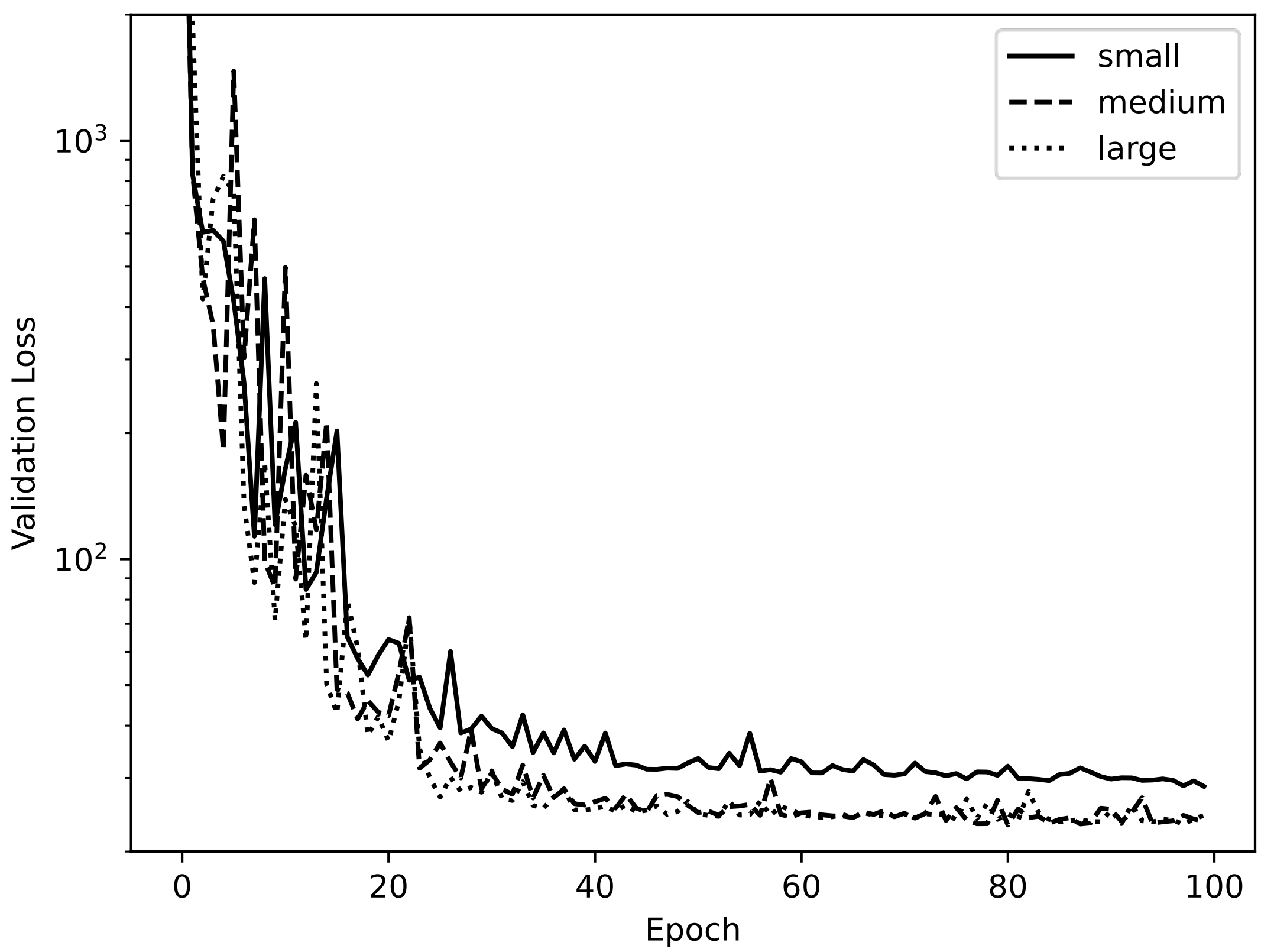

**Figure 2.** Validation loss during training for the three Nutmeg models. All models were mostly converged by 50 epochs, with only small improvements after that point. None of them showed signs of increasing validation loss, which would suggest overfitting.

# Test Set

The validation set consists of randomly chosen conformations from the SPICE 2 dataset. It provides information about how well models generalize to novel conformations of molecules on which they were trained. We also want to evaluate how well they generalize to novel molecules that were not present in the training set. In addition, we want to test the hypothesis that a model trained only on small molecules can scale to larger ones.

We therefore created a test set consisting of molecules not present in the training set. It includes the following.

- 200 randomly chosen molecules from Ligand Expo with between 40 and 50 atoms. The SPICE ligand amino acid pair subset includes molecules from Ligand Expo, but only ones with up to 36 atoms, so the training set does not include these molecules. It does include many PubChem molecules in the same size range, however. These molecules are used to test how well the models can generalize to new molecules of the same size as the training molecules.
- 200 randomly chosen molecules from Ligand Expo with between 70 and 80 atoms. These are larger than any single molecule found in the SPICE dataset, although some dimers and solvated molecules contain this many atoms in total. These molecules help to evaluate how well models can scale to larger molecules than they were trained on.
- 200 pentapeptides. The amino acids were randomly chosen from among the 20 natural ones, and each peptide was capped with ACE and NME groups. They cover a range of sizes, with the very largest having 110 atoms. The training set contains all possible dipeptides, so these molecules test whether the models can scale to longer peptides.
- 200 dimers. Each consists of a different molecule from Ligand Expo with between 15 and 36 atoms, paired with a single amino acid. These dimers consist of molecules that individually are present in the training set, but interacting non-covalently in pairs that are not present in the training set. They test how well models can generalize to novel non-covalent interactions.

10 conformations were generated for each molecule. For the Ligand Expo molecules, RDKit was used to generate 10 starting conformations, and each was followed by 5 ps of molecular dynamics at 300K with the GFN-FF force field. For the peptides, 500 ps of molecular dynamics at 300K with the Amber ff14 force field[33] was performed, and a configuration was saved every 50 ps. Dimers were generated using the same procedure described for the SPICE ligand amino acid subset, but excluding any pair present in the training set and drawing conformations from multiple PDB files to get 10 conformations for each dimer. Forces and energies were computed with Psi4 1.8.2 at the same level of electronic structure theory used for the SPICE dataset. The electronic structure calculations failed to converge for 20 conformations, and another four were excluded due to the presence of forces above the cutoff of 1 hartree/bohr, giving a total of 7976 conformations in the test set.

# Results

## Validation Set Accuracy

We first evaluate the performance of the three Nutmeg models on the validation set. This provides information about their accuracy on novel conformations of molecules on which they were trained. Table 3 shows the distribution of absolute errors of each model.

| | Small | Medium | Large |
|---|---|---|---|
| Mean | 2.60 | 2.19 | 1.85 |
| 50th percentile | 1.71 | 1.50 | 1.28 |
| 90th percentile | 5.67 | 4.68 | 3.87 |
| 99th percentile | 13.54 | 10.66 | 9.24 |

**Table 3.** The distribution of absolute errors on the validation set in kJ/mol for the three Nutmeg models.

All three models have mean absolute errors (MAE) much less than 4.184 kJ/mol (1 kcal/mol), a number often cited as a standard for chemical accuracy.[34] For all three models, the errors are distributed very non-uniformly. The median (50th percentile) absolute error is much lower than the mean, while a tiny fraction of conformations have much larger errors that bias the mean upward. The highest 1% of errors are at least seven times higher than the median for all models.

A major factor contributing to the variation in errors is the size of a molecule or cluster. Figure 3 shows the MAE as a function of the number of atoms. All three models have remarkably similar results, not only in the overall shape but even in many of the details, with the larger models being shifted to lower error.

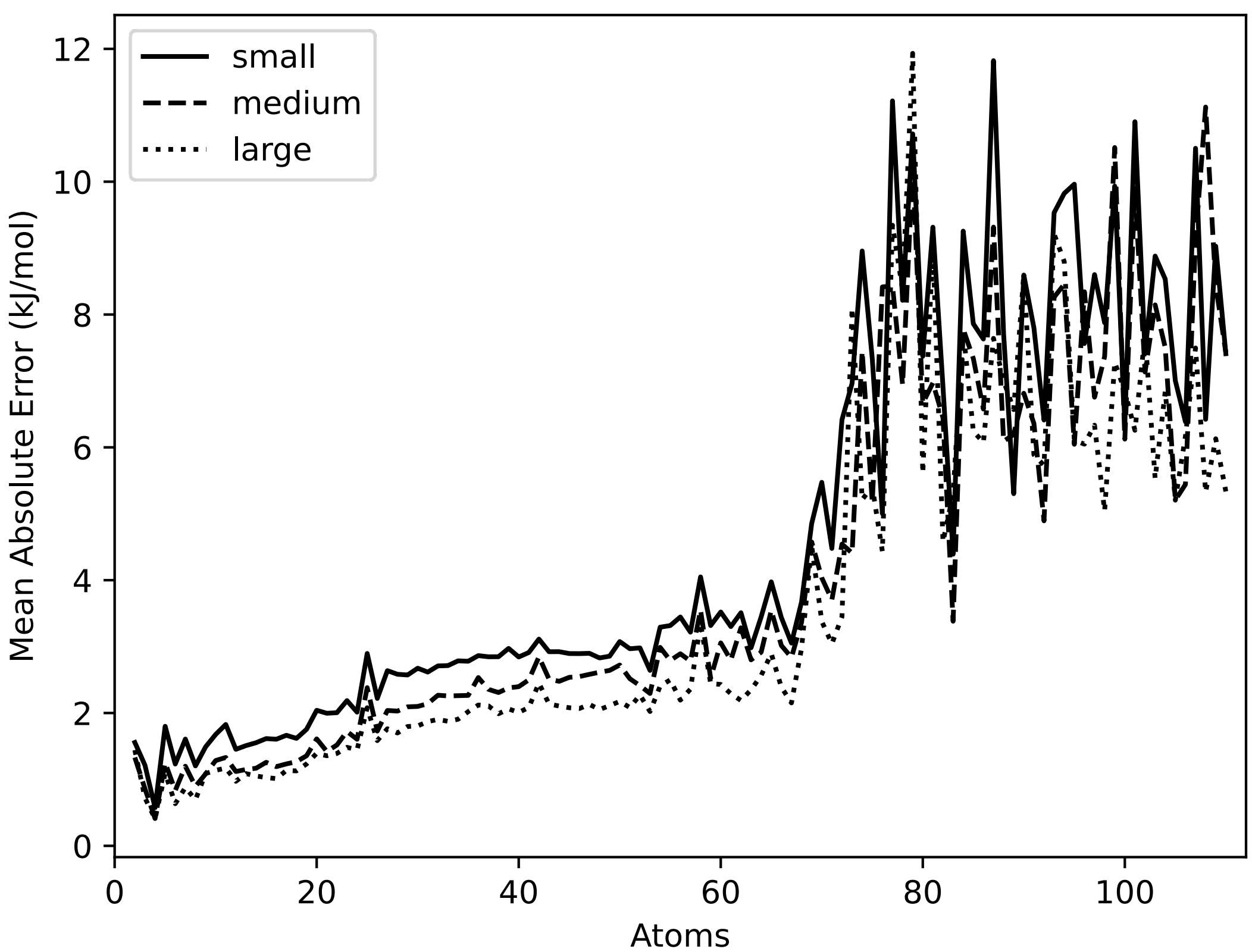

**Figure 3.** Mean absolute error on the validation set as a function of the number of atoms in a molecule or cluster for the three Nutmeg models.

Up to 70 atoms, there is a gradual increase in error with size, but all three models remain quite accurate. Above this point there is a rapid jump, with the mean error roughly doubling between 70 and 80 atoms. Beyond 80 atoms there is no obvious further increase, and the MAE remains flat up to 110 atoms. The fluctuations in the graph are also much larger beyond this point, but this might only reflect that there are many fewer samples for larger numbers of atoms.

The jump in error appears to be caused by the finite cutoff of the message passing operations, which prevents them from exchanging information between distant parts of a large molecule. To test this, we trained another model with identical hyperparameters to Nutmeg-small, except reducing the cutoff distance from 1.0 nm to 0.6 nm. This shorter cutoff is more typical of those used in other published models. [26,35] The effect is shown in Figure 4. For sufficiently small molecules it has a minimal effect on accuracy, but for larger molecules it has a large effect. The jump in error now occurs at around 50 atoms, compared to 70 atoms for the model with the longer cutoff.

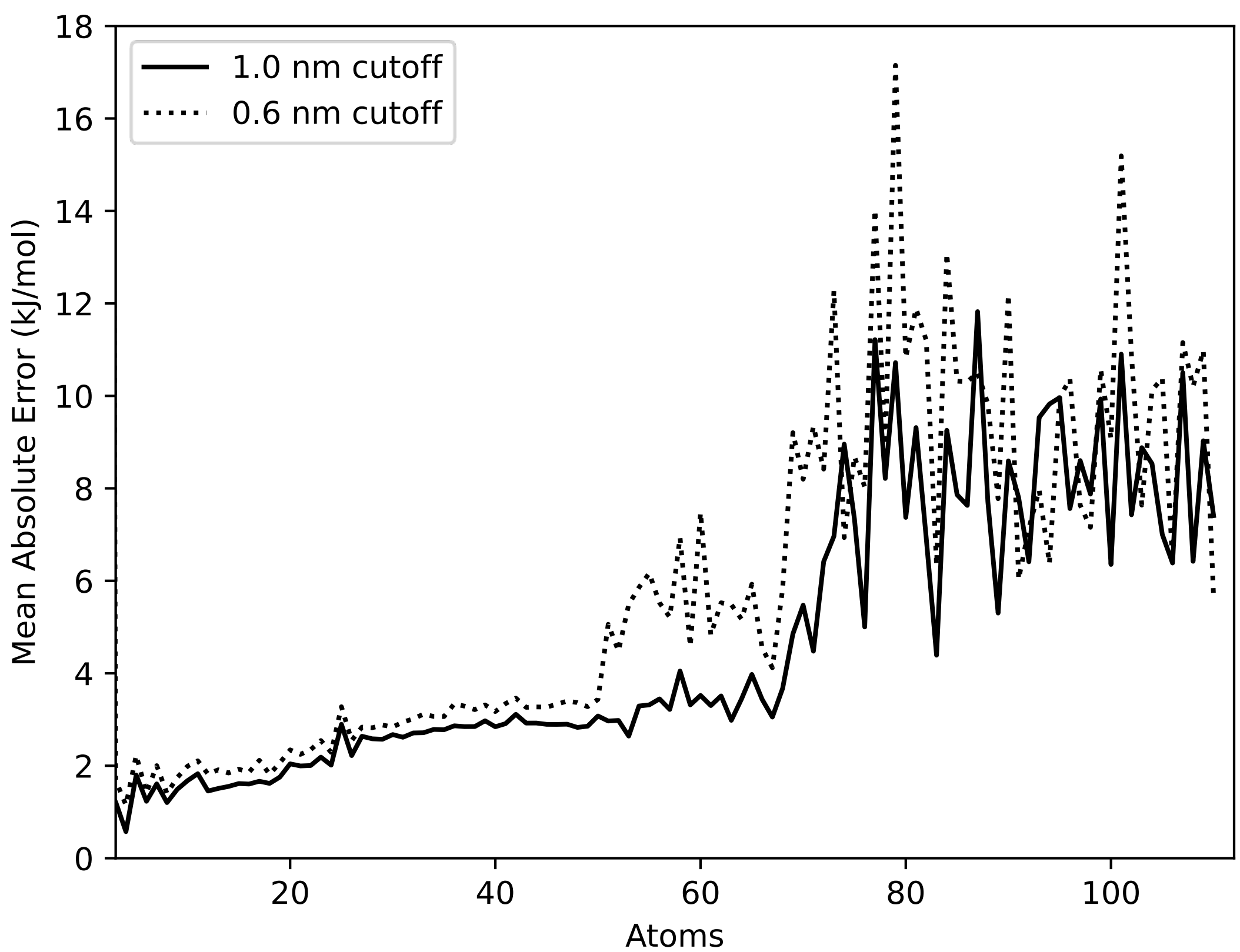

**Figure 4.** Accuracy of two models that use identical hyperparameters except for the cutoff distance on message passing. Reducing the cutoff has minimal effect on sufficiently small molecules, but the jump in error is shifted to lower numbers of atoms.

We conclude that it is important to evaluate models on molecules of a variety of sizes. Accuracy on small molecules does not necessarily imply accuracy on larger ones, and abrupt changes in behavior can occur as the number of atoms increases.

## Test Set Accuracy

We now examine accuracy on the test set, which measures how well the models generalize to new molecules on which they were not trained. Table 4 shows the mean and median absolute errors on the four groups of molecules that make up the test set: small ligands (40-50 atoms), large ligands (70-80 atoms), pentapeptides (68-110 atoms), and dimers (34-72 atoms).

| | | **Small** | **Medium** | **Large** |
|---|---|---|---|---|
| Small Ligands | Mean | 7.32 | 6.87 | 6.52 |
| | Median | 3.20 | 2.81 | 2.56 |
| Large Ligands | Mean | 8.17 | 7.68 | 6.70 |
| | Median | 4.48 | 4.44 | 3.35 |
| Peptides | Mean | 14.77 | 13.32 | 12.14 |
| | Median | 7.94 | 5.82 | 4.80 |
| Dimers | Mean | 11.82 | 11.86 | 11.36 |
| | Median | 2.24 | 2.12 | 1.77 |

**Table 4.** Absolute errors in kJ/mol on the test set.

Comparing to validation set samples of similar size as shown in Figure 3, we see that the test set errors are somewhat higher. Other patterns are as one expects: larger molecules tend to have larger errors; larger models have better accuracy; and median errors are always much lower than mean errors, indicating that a small number of samples with high errors strongly influence the mean. This is especially pronounced for the dimers, which have the largest difference between mean and median of any group. Their median errors are the lowest of any group, yet the mean errors are higher than any group except peptides.

In many cases one is not interested in absolute energies, only in the energy differences between conformations. We can measure this by subtracting off the mean energy over the ten conformations for each molecule. The resulting errors are shown in Table 5. Many of the values are greatly reduced. The effect is especially dramatic for the pentapeptides, with the MAE reduced by a factor of 2.7, 2.8, and 3.0 for the small, medium, and large models respectively. In fact, the pentapeptides now have lower MAE than the large ligands, despite being larger molecules. We hypothesize this reflects the fact that the training set contains data on dipeptides, so the longer peptides in the test set do not involve any new areas of chemical space. When the models are applied to molecules that are larger but chemically similar to the training data, there is an overall shift compared to the true absolute energies, but the energy differences between conformations remain accurate.

| | | Small | Medium | Large |
|---|---|---|---|---|
| Small Ligands | Mean | 4.56 | 3.94 | 3.51 |
| | Median | 2.50 | 2.06 | 1.69 |
| Large Ligands | Mean | 7.09 | 6.47 | 5.69 |
| | Median | 3.69 | 3.24 | 2.70 |
| Peptides | Mean | 5.47 | 4.71 | 3.99 |
| | Median | 3.96 | 3.37 | 2.94 |
| Dimers | Mean | 9.68 | 9.89 | 9.86 |
| | Median | 1.87 | 1.58 | 1.35 |

**Table 5.** Errors in conformational energies. Values are the absolute errors in kJ/mol on the test set after subtracting off the mean energy of the ten conformations for each molecule.

Another way to analyze this is to directly compute the differences in energy between conformations for the same molecule. The result of this analysis is shown in Figure 5 for Nutmeg-large. For each molecule, it shows the mean absolute error in energy differences versus the magnitude of the energy differences. The two are strongly correlated: larger energy differences tend to have proportionally larger absolute errors. The errors in energy differences are an order of magnitude smaller than the differences themselves, indicating that the model does an excellent job of ranking conformations by energy.

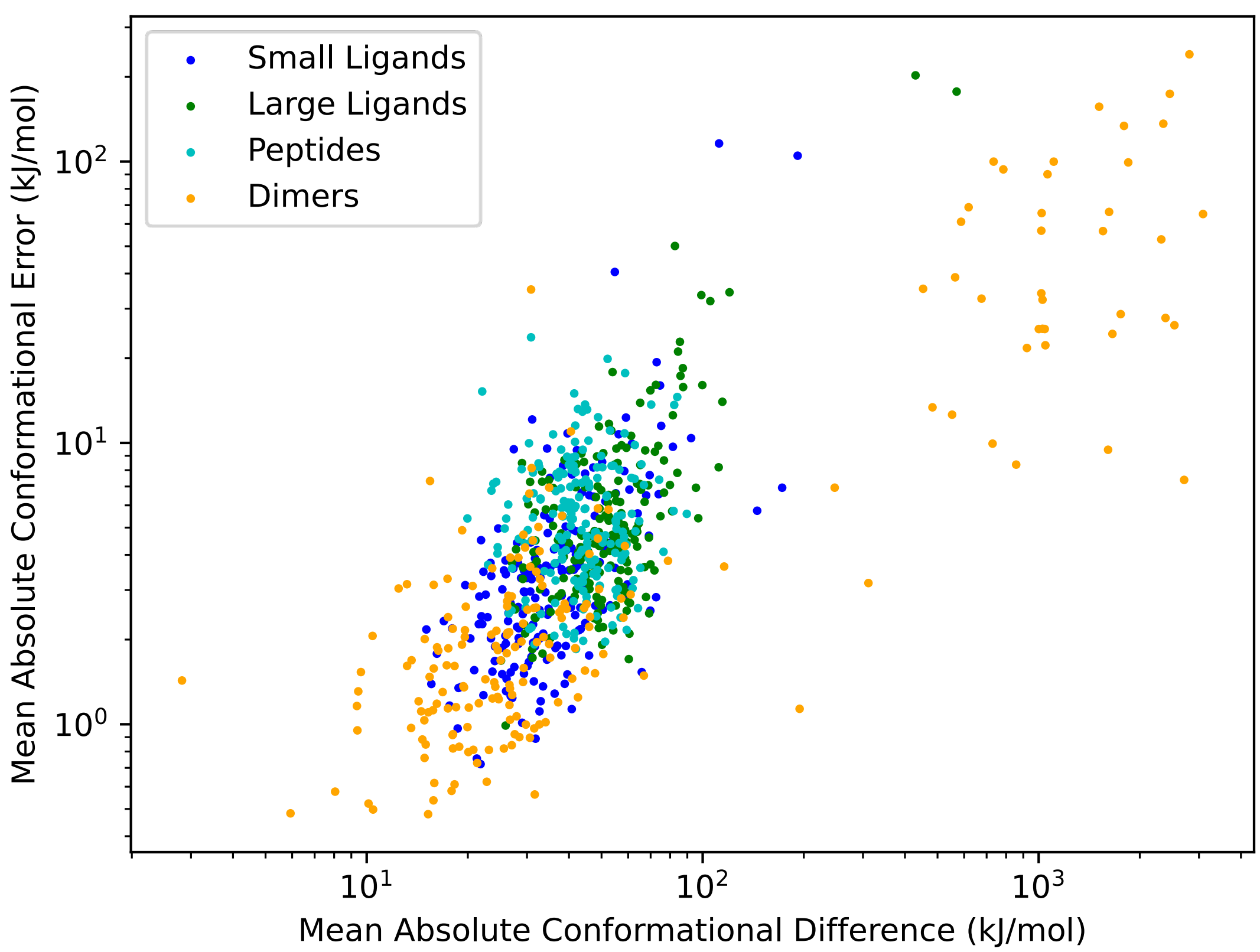

**Figure 5.** Accuracy of energy differences between conformations for Nutmeg-large. There is one point for every molecule or dimer in the test set. Horizontal axis: the mean absolute difference in reference energy over all pairs of conformations. Vertical axis: the mean absolute error in the predicted energy difference over all pairs of conformations.

The dimers are noteworthy, since they include both the lowest and the highest energy differences. This explains why the mean error is so much higher than the median error for this group. A fraction of dimers have exceptionally large energy differences between conformations, and correspondingly larger errors that increase the mean. This does not mean the model is less accurate for dimers. In fact, the dimers with very large energy differences deviate from the trend of the other molecules, actually having lower error than one would expect.

Another question to answer is whether our method of injecting precomputed partial charges into the model is effective for simulating charged and polar molecules. Figure 6 shows the MAE as a function of the number of atoms with a nonzero partial charge, both with and without subtracting per-molecule mean energies. The left panel shows the error in absolute energies. It shows a dramatic increase in the error for molecules with charged groups, reaching hundreds of kJ/mol for ones with four charged atoms. The right panel shows the same data, but after subtracting off the mean energy of the ten conformations for each molecule. The pattern is now completely different, with very little change in the error as the number of charged groups increases. The remaining variation still apparent may just reflect the fact that larger molecules tend to have more charges. The molecules with three charged groups have an average of 94 atoms, compared to only 65 atoms for the ones with no charges. This alone could easily account for the difference in error seen in the graph. We conclude that the models produce an overall shift in the absolute energies of charged and polar molecules, but energy differences between conformations are still quite accurate.

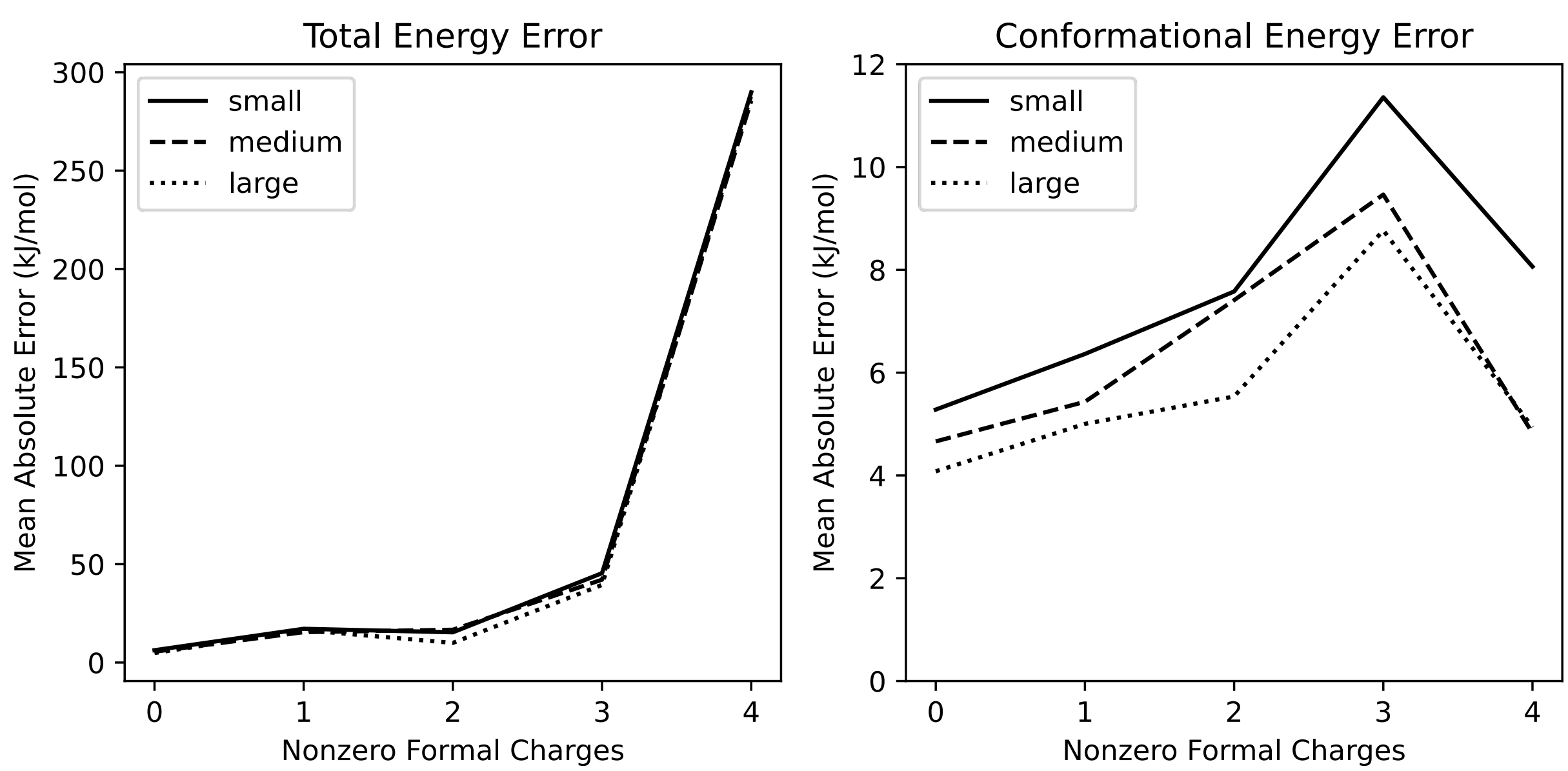

**Figure 6.** Mean absolute error on the test set as a function of the number of atoms with nonzero formal charge. Left: error in the absolute energy. Right: error after subtracting off the mean energy of all conformations for each molecule.

To evaluate how important the partial charges are for achieving this accuracy, we trained two additional models with identical hyperparameters to Nutmeg-small, except that one injects formal charges into the embedding layer in place of partial charges, and one does not inject any charges at all. The effect is shown in Figure 7. For molecules with no charged groups, the model that does not inject charges is only slightly less accurate (MAE has increased from 5.3 to 6.7 kJ/mol), but for molecules with charges the accuracy is dramatically worse. For the molecules with 2 or 3 nonzero formal charges, omitting the partial charges increases the MAE by more than a factor of four. The model that uses formal charges avoids this catastrophic loss of accuracy for charged molecules, but it is consistently less accurate than using partial charges, regardless of the charge of the molecule.

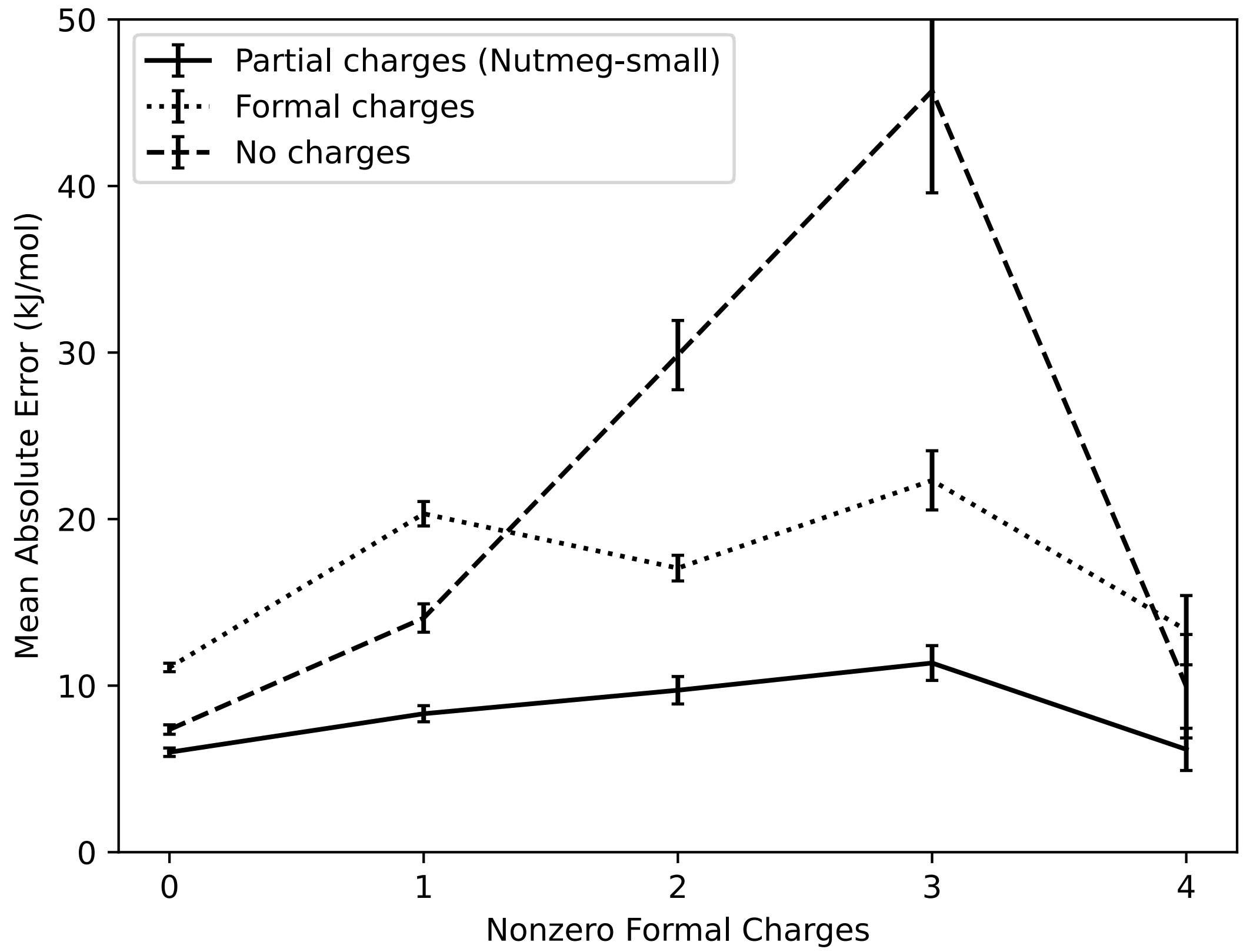

**Figure 7.** Mean absolute error on the test set after subtracting off the mean energy of all conformations for each molecule. Error bars correspond to one standard error. The three models have identical hyperparameters, differing only in what information is injected into the embedding layer: Gasteiger partial charges, formal charges, or no charges at all. The model that uses formal charges has consistently higher error than the one that uses partial charges, while the model that does not use any charges at all has dramatically higher error on molecules containing charged groups.

# Simulation Stability

Having low error on test conformations is a necessary but not sufficient condition for a potential energy function to be useful for running simulations. We now evaluate whether the Nutmeg models are able to produce stable simulations.

There are two distinct types of instability that must be considered. The first is problems that cause the integration algorithm to become unstable. This can happen, for example, if forces are too large or change too rapidly. Mild cases of instability lead to integration errors that increase the simulation temperature. More severe cases lead to abrupt failures in which molecules explode or coordinates take on huge values.

Another case is that the integration algorithm is stable, but the molecule itself is not stable. This can happen, for example, if barriers to bond breaking are unrealistically low, allowing the atoms that form a molecule to come apart.

To evaluate simulation stability, we performed a series of 1800 short molecular dynamics simulations: one for each of the 600 molecules in the test set and each of the three Nutmeg models. Beginning from the first conformation in the test set, each molecule was first energy minimized, then simulated at 300K with a 1 fs time step for 10,000 steps (10 ps). The total system kinetic energy was saved every 100 fs. Simulations in this section and the following sections were performed with OpenMM 8.1.1.

Each simulation was checked for signs of instability as follows.

- No bonds should have broken. Specifically, the length of each bond in the final conformation should be no more than 0.5 Å greater than in the original conformation.
- No component of the force on any atom should be larger than 5000 kJ/mol/nm in the final conformation. This is not a strict requirement for stability, but rather an approximate threshold of the largest forces typically encountered in room temperature molecular simulations.
- The average kinetic energy over the last 5 ps should correspond to a temperature below 400K. Given the small sizes of the molecules and short simulation lengths, large fluctuations in the kinetic energy are to be expected, so a stricter cutoff is not possible.

Among the 1800 simulations, no instances of kinetic energies exceeding the temperature threshold were observed. There were three instances of forces greater than 5000 kJ/mol/nm, but only by a small amount (all of them less than 6000 kJ/mol/nm), and they do not seem to have caused problems for stability.

One molecule (Ligand Expo ID 1R3) was found to be unstable with all three models. It contains a hypervalent sulfur bonded to two oxygens and two carbons. During the initial energy minimization, one of the oxygens detaches and remains separate through the simulation. No other instances of broken bonds occurred for the small or large models. For the medium model, two additional instances of broken bonds occurred (Ligand Expo IDs 7CQ and NXQ), both of them also involving hypervalent sulfurs. Unlike the broken bond in 1R3, which occurs during the initial energy minimization and is fully reproducible, these two appear to be stochastic. When the simulations of them were repeated, no broken bonds were observed.

To understand the cause of the instability, we energy minimized molecule 1R3 at the ωB97M-D3(BJ)/def2-TZVPPD level of theory. ASE 3.22.1[36] and Psi4 1.8.2 were used to perform 30 steps of L-BFGS minimization. Over the course of the minimization, the sulfur-oxygen bond stretched from 1.67 Å in the initial conformation to 2.27 Å at the end. This suggests the molecule really is unstable, at least at the level of theory used to generate the training data, and the Nutmeg models have correctly learned the instability.

# Simulation Speed

Another requirement for a model to be useful is that it can be computed fast enough to produce simulations of useful length. To evaluate this, we measured the simulation speed of three molecules drawn from the training and test sets with 25, 50, and 100 atoms (PubChem substance IDs 135001062 and 103939106, and the peptide with sequence Trp-Gly-Tyr-Lys-Pro). Each was simulated on an NVIDIA RTX 4080 GPU with each of the three models and a 1 fs time step. The speeds are shown in Table 6, both as the time in milliseconds (ms) to perform a single time step and as the overall speed in ns/day.

| | **Step Time (ms)** | | | **Speed (ns/day)** | | |
|---|---|---|---|---|---|---|
| **Atoms** | **Small** | **Medium** | **Large** | **Small** | **Medium** | **Large** |
| 25 | 6.5 | 6.5 | 9.0 | 13.30 | 13.23 | 9.60 |
| 50 | 6.9 | 9.1 | 13.8 | 12.49 | 9.54 | 6.27 |
| 100 | 13.3 | 23.8 | 38.0 | 6.50 | 3.63 | 2.28 |

**Table 6.** The speed of the three Nutmeg models simulating molecules with 25, 50, or 100 atoms. Simulations used a 1 fs time step and were run on an NVIDIA RTX 4080 GPU.

A bottleneck to faster performance on small molecules is the difficulty of fully utilizing the GPU. While simulating the smallest molecule with the smallest model, GPU utilization is reported as only about 50%. This leads to only a modest increase in cost with model size. For the 25 atom molecule, the small and medium models are nearly indistinguishable in speed, and the large model is only 38% slower. For the larger molecules the differences become more pronounced, with the large model being 2.9x slower than the small one on the 100 atom molecule.

We are hopeful the performance can be improved in the future. Two promising mechanisms are CUDA graphs, which reduce the overhead of launching individual GPU kernels, and the new compilation mechanism introduced in PyTorch 2, which produces better optimized GPU kernels. Taking advantage of either of these mechanisms will require code changes to PhysicsML.

We can put these speeds into perspective by comparing to other methods of computing molecular forces. When we performed the DFT calculations for the 100 atom peptide with Psi4, a 32 core CPU required approximately three hours for each conformation. The Nutmeg models require only a few ms to compute forces, indicating a speedup of almost six orders of magnitude. On the other hand, they are orders of magnitude slower than classical point charge force fields. For comparison, using OpenMM with the Amber ff14[33] force field on the same RTX 4080 GPU, one can achieve over 500 ns/day for a system with 100,000 atoms, three orders of magnitude larger than the simple peptide.

We view Nutmeg and other similar models as a faster alternative to quantum chemistry methods rather than as a more accurate alternative to force fields. They are useful for performing accurate calculations on small molecules, but at least in the near term will probably remain too slow to be practical for simulating larger systems such as solvated proteins.

## Torsion Profiles

Dihedral rotations are the primary source of large-scale conformational changes in molecules. To produce an accurate ensemble of conformations, a potential function must be able to reproduce the energy profiles of dihedral torsion angles. Ideally, it should reproduce both the positions of the energy minima and the heights of the barriers between them. We expect the latter to be a particularly challenging test for these models. Barrier heights are determined by high energy transition states that are rarely visited in equilibrium. Most of the conformations in the SPICE dataset are drawn from equilibrium thermal distributions, so the transition states are likely to be poorly sampled.

To evaluate the accuracy of torsion profiles, we use the TorsionNet 500 benchmark dataset.[37] It is a collection of torsion profiles for 500 diverse molecules. Each molecule has 24 conformations sampling a full rotation about the central bond, and corresponding energies computed at the B3LYP/6-31G** level of theory. This is less accurate than the level of theory used by SPICE, so we recomputed the energies of all conformations at the ωB97M-D3(BJ)/def2-TZVPPD level of theory to make the results directly comparable.

Figure 8 shows a set of representative torsion energy profiles. For all models, the agreement with the DFT reference values is excellent. Table 7 shows the mean signed and unsigned errors in barrier heights, defined as the difference between the highest and lowest energy points on a profile. For comparison, the thermal energy kT at 300K is 2.49 kJ/mol. Typical errors are less than this, suggesting they should have little impact on transition rates for dihedral rotations.

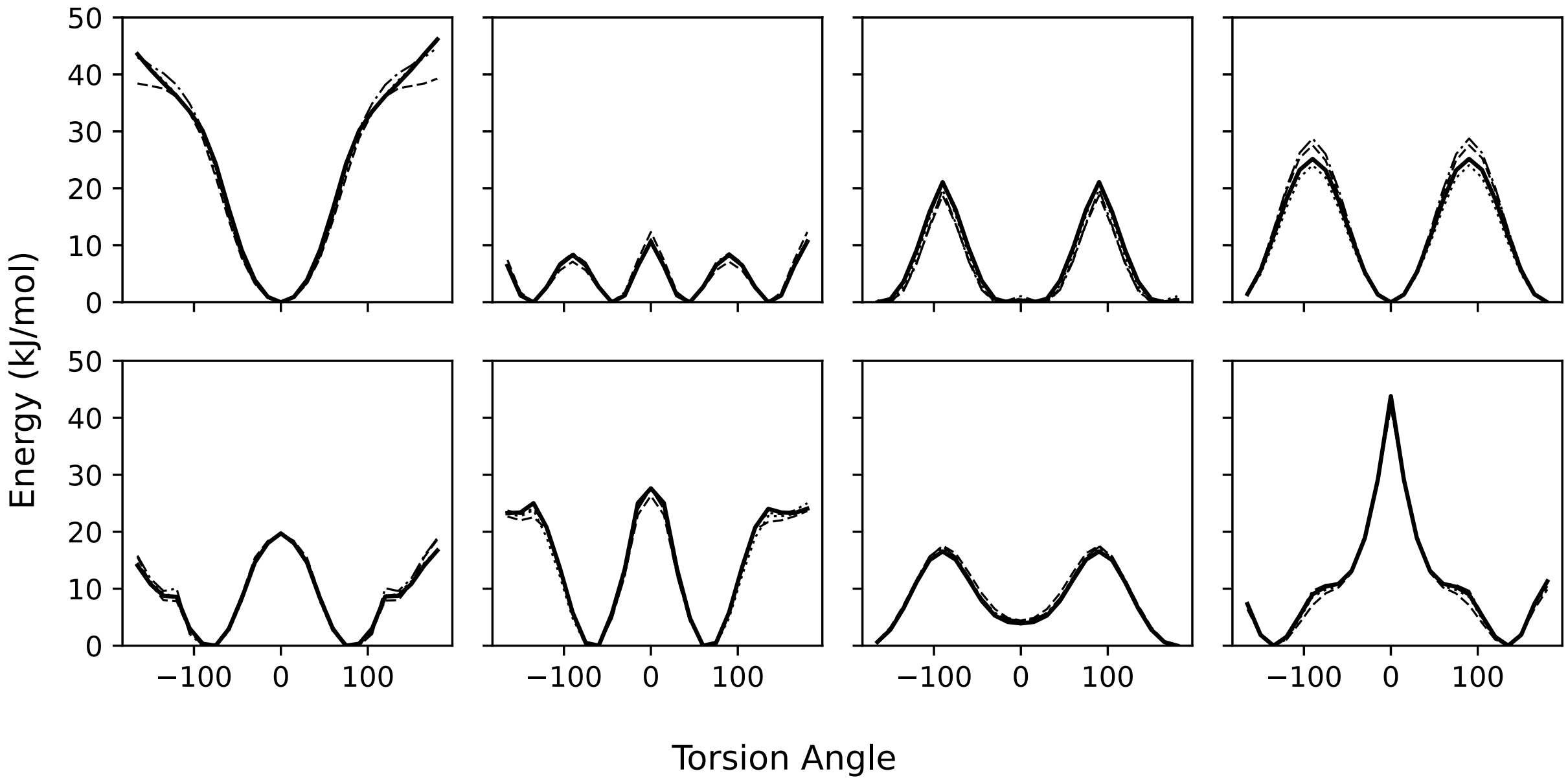

**Figure 8.** Energy profiles for eight representative torsions from the TorsionNet 500 dataset. The solid line is the DFT reference energies, while the other lines are for the small (dashed), medium (dot-dash) and large (dotted) Nutmeg models. To make differences in barrier heights more visible, all curves are shifted to place the minimum energy at 0 kJ/mol.

| | Small | Medium | Large |
|---|---|---|---|
| Mean Absolute Error | 1.69 | 1.30 | 1.06 |
| Mean Signed Error | -0.32 | -0.10 | -0.02 |

**Table 7.** Mean errors in kJ/mol in barrier heights over the molecules in the TorsionNet 500 dataset compared to the ωB97M-D3(BJ)/def2-TZVPPD reference energies.

# Bulk Water

The SPICE dataset includes clusters with up to 30 water molecules. We now examine whether this is sufficient to let it accurately simulate bulk water. We used Nutmeg-small to simulate a 2.2 nm cube of water containing 346 molecules (1038 atoms). This is an order of magnitude larger than anything in the training set. It also is qualitatively different from any of the training data, being a bulk material in a periodic box rather than a single isolated cluster. Both of these make it a particularly challenging test. The system was simulated at 300K with a 1 fs time step for 200 ps. The oxygen-oxygen radial distribution function is shown in Figure 9. For comparison, we perform similar simulations with three conventional water models: one fixed charge model (TIP4P-FB[17]) and two polarizable models (AMOEBA[18] and SWM4-NDP[38]).

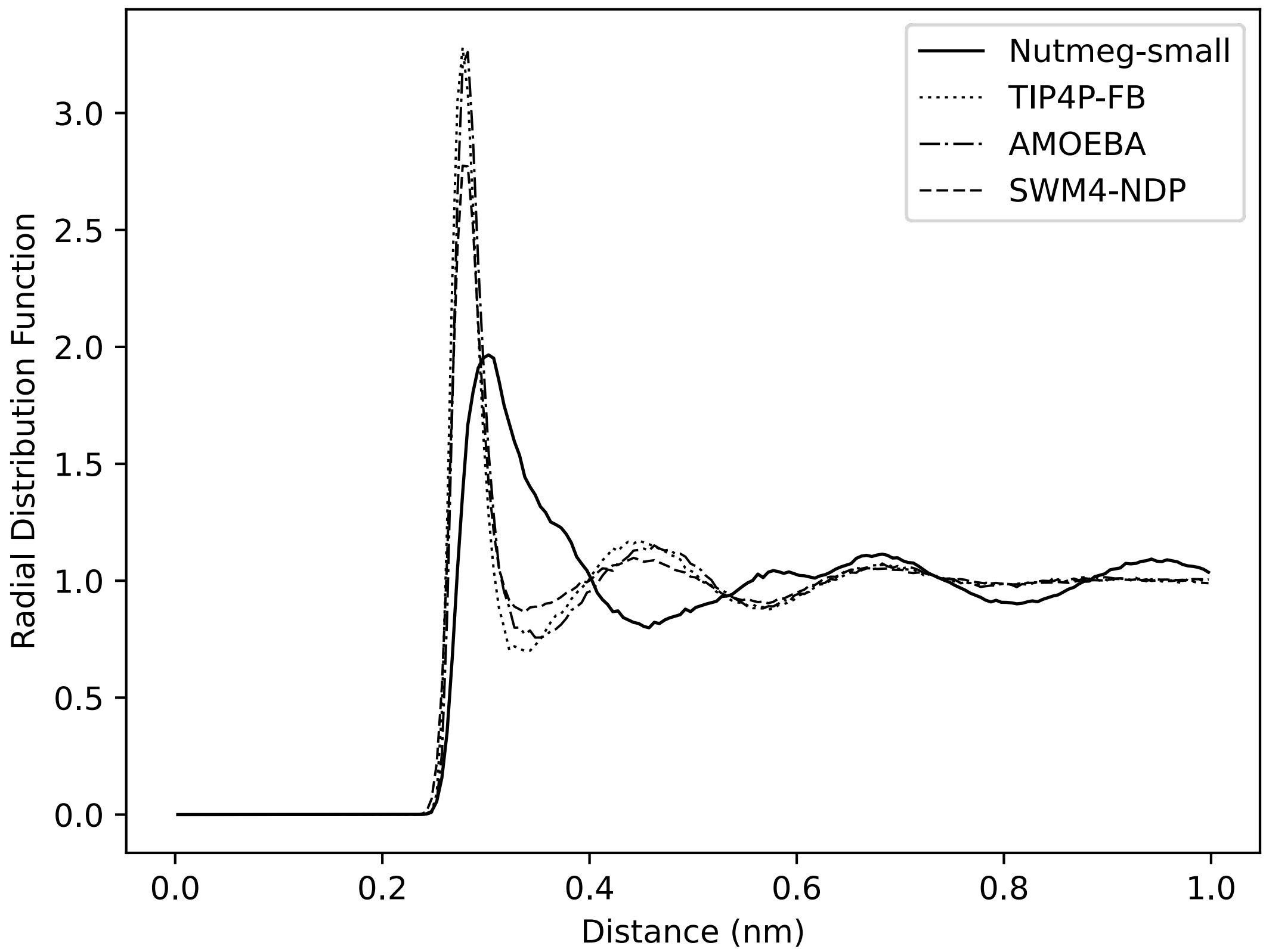

**Figure 9.** Oxygen-oxygen radial distribution function for bulk water from simulations with Nutmeg-small and three conventional water models.

We see that Nutmeg-small does a poor job of simulating bulk water. All of the peaks are shifted to too large values, with significant structure being apparent all the way out to 1 nm. We therefore do not recommend the use of Nutmeg models for simulating bulk materials. Understanding the reason for this and identifying how to improve it is an important subject for future research. Possibly the clusters of 30 water molecules were too small for the model to accurately learn the behavior of bulk water. Good results have been reported simulating water with the MACE-OFF23 models, which were trained on a similar dataset but including larger clusters with up to 50 molecules.[35]

# Conclusions

SPICE 2 is a major expansion and improvement to the dataset. It greatly expands the sampling of chemical space and includes more data on non-covalent interactions. It is a useful resource for anyone creating machine learning potentials.

The Nutmeg models are a powerful tool for simulating drug-like small molecules, with three models of different sizes to give a choice of tradeoffs between speed and accuracy. They support a wide range of chemical space, including 17 elements and charged molecules.

For nonpolar molecules with up to 70 atoms, all three models do an excellent job of reproducing absolute formation energies. For larger molecules there is a decrease in accuracy, although in some cases this seems to mostly consist of a constant shift to the energy, with energy differences between conformations still being accurate. Similarly, the presence of charged groups in a molecule leads to a global shift in the energy, but has little effect on energy differences between conformations.

The models do an excellent job of reproducing torsion energy profiles, even though no effort was made to sample transition states in the training set. The details of the profiles are accurately reproduced, and errors in barrier heights are small compared to thermal energy at room temperature.

The accuracy of the Nutmeg models on individual molecules does not extend to bulk materials. Water structure is found to be poorly reproduced. We therefore recommend using the Nutmeg models only for individual molecules or small clusters.

The Nutmeg models are not intended to simulate chemical reactions. Any change to bonding patterns would invalidate the atomic partial charges. For this reason, it is important to monitor simulations for broken bonds. This is expected to be an infrequent occurrence and not to present a major difficulty. All instances of broken bonds encountered during testing involved sulfur atoms bonded to three or more other atoms. This appears to be a particularly challenging motif for the models, and extra care should be used when simulating molecules that contain it.

# Availability

SPICE version 2 is available from Zenodo at https://doi.org/10.5281/zenodo.10975225. The scripts and data files used in generating it are available from GitHub at https://github.com/openmm/spice-dataset. The Nutmeg models are available from GitHub at https://github.com/openmm/nutmeg. The test set is available from Zenodo at https://doi.org/10.5281/zenodo.11455132.

# Acknowledgments

Research reported in this publication was supported by the National Institute of General Medical Sciences of the National Institutes of Health under award number R01GM140090 (TEM, PE) and award number R35 GM152017 (JDC). This research was funded in part through the NIH/NCI Cancer Center Support Grant P30 CA008748 (JDC). JDC acknowledges financial support from the Sloan Kettering Institute. BPP acknowledges support from the National

Science Foundation via grant CHE-2136142. Some of the computing for this project was performed on the Sherlock cluster. We would like to thank Stanford University and the Stanford Research Computing Center for providing computational resources and support that contributed to these research results.

We thank Yuezhi Mao for his assistance in investigating the unstable molecules. We thank Joshua Rackers, Geoffrey Wood, and Pavan Behara for their suggestions related to the design of the dataset.

# Disclaimer

The content is solely the responsibility of the authors and does not necessarily represent the official views of the National Institutes of Health.

# Disclosures

JDC is a current member of the Scientific Advisory Board of OpenEye Scientific Software, Redesign Science, Ventus Therapeutics, and Interline Therapeutics, and has equity interests in Redesign Science and Interline Therapeutics. The Chodera laboratory receives or has received funding from multiple sources, including the National Institutes of Health, the National Science Foundation, the Parker Institute for Cancer Immunotherapy, Relay Therapeutics, Entasis Therapeutics, Silicon Therapeutics, EMD Serono (Merck KGaA), AstraZeneca, Vir Biotechnology, Bayer, XtalPi, Interline Therapeutics, the Molecular Sciences Software Institute, the Starr Cancer Consortium, the Open Force Field Consortium, Cycle for Survival, a Louis V. Gerstner Young Investigator Award, and the Sloan Kettering Institute.

A complete funding history for the Chodera lab can be found at http://choderalab.org/funding.